\documentclass[aps,prb,twocolumn,superscriptaddress,showpacs]{revtex4}

\usepackage{amsmath}
\usepackage{bm}
\usepackage{graphicx}

\begin{document}

\title{Transport through a quantum wire with a side quantum-dot array}

\author{P.\ A.\ Orellana}
\affiliation{Departamento de F\'{\i }sica, Universidad Cat\'{o}lica del Norte,
Casilla 1280, Antofagasta, Chile}

\author{F.\ Dom\'{\i}nguez-Adame}
\affiliation{Departamento de F\'{\i}sica de Materiales, Universidad
Complutense, E-28040 Madrid, Spain}

\author{I.\ G\'{o}mez}
\affiliation{Departamento de F\'{\i}sica de Materiales, Universidad
Complutense, E-28040 Madrid, Spain}

\author{M.\ L.\ Ladr\'{o}n de Guevara}
\affiliation{Departamento de F\'{i}sica, P. Universidad Cat\'{o}lica de
Chile, Casilla 306, Santiago 22, Chile.}

\date{\today}

\begin{abstract}

A noninteracting quantum-dot array side-coupled to a quantum wire is studied.
Transport through the quantum wire is investigated by using a noninteracting
Anderson tunneling Hamiltonian. The conductance at zero temperature develops an
oscillating band with resonances and antiresonances due to constructive and
destructive interference in the ballistic channel, respectively. Moreover, we
have found an odd-even parity in the system, whose conductance vanishes for an
odd number of quantum dots while becomes $2e^2/h$ for an even number. 
We established an explicit relation between this odd-even parity, and the
positions of the resonances and antiresonances of the conductivity with the spectrum
of the isolated QD array

\end{abstract}


\maketitle

\section{Introduction}

Recent progress in nanofabrication of quantum devices enables to study
electron transport through quantum dots (QDs) in a very controllable way.%
\cite{qd1,qd2} QDs are very promising systems due to their physical
properties as well as their potential application in electronic devices.
These structures are small semiconductor or metal structures in which
electrons are confined in all spatial dimensions. As a consequence,
discreteness of energy and charge arise. For this reason QDs are often
referred  as \emph{artificial atoms}. In contrast to real atoms, different
regimes can be studied by continuously changing the applied external
potential.

If the single QDs is referred as \emph{artificial atoms} a QD array can be
considered as \emph{artificial molecule} or \emph{artificial crystal}.\cite
{dqd1,dqd2,qdaN} Latest advances in nanotechnology make it possible to
fabricate QD arrays. In linear QD arrays leads are attached to their ends
and the current through them is measured while external parameters such a
gated voltage, magnetic field, and temperature are varied. In resonant
tunneling regime, the electronic transport through QD array becomes
sensitive to precise matching of the electron levels in the dots that can be
controlled experimentally. For other hand, a linear QD array can be seen as
an one dimensional chain of sites. This type of chain coupled to the
continuum states shows an even-odd parity effect in the conductance when the
Fermi energy is localized in the center of the energy band.\cite{oguri,zeng}
The conductance is $2e^{2}/h$ for odd samples and is smaller for even parity.%
\cite{oguri,zeng}

The aim of this work is to study theoretically the transport properties of
an alternative configuration of a side-coupled QD array attached to a
perfect quantum wire (QW). In this case the QD array acts as scattering
center for transmission through the QW. This configuration can be regarded
as a quantum wave guide with side-stub structures, similar to those reported
in Refs.~\onlinecite{stub1,stub2,stub3}. In contrast to the embedded QD
array, the transmission through the side-coupled QD array consists of the
interference between the ballistic channel and the resonant channels
from the QD array. For a uniform side QD array, we found that the system develops an oscillating 
band with resonances ( perfect transmission) and antiresonances (perfect
reflection). In addition, we found an odd-even parity of the number of QDs
in the array, namely perfect transmission takes place if this number is even
($G=2e^{2}/h$) but perfect reflection arises for an odd number ($G=0$). This
result is opposed to that found in embedded QD arrays. \cite{oguri,zeng}
We established an explicit relation between this odd-even parity, and the
positions of the resonances and antiresonances of the conductivity with the spectrum
of the isolated QD array.

\section{Model}

\begin{figure}[th]
\centerline{\includegraphics[width=50mm]{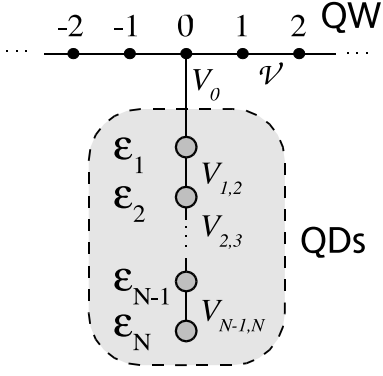}}
\caption{Side-coupled quantum dot array attached to a perfect quantum wire.}
\label{fig1}
\end{figure}

Let us consider a QW with a side-coupled QD array. The array consists of $N$
QDs connected in a series by tunnel coupling, as shown in Fig.~\ref{fig1}.
The system is modeled by using a noninteracting Anderson tunneling
Hamiltonian\cite{qdaN} that can be written as 

\begin{equation} H=H_{\text{QW}}+H_{\text{QD}}^{N}+H_{\text{QD-QW}},
\end{equation}
with
\begin{eqnarray}
&&H_{\text{QW}} =v\sum_{i\neq j}\,c_{i}^{\dagger }c_{j},  \nonumber \\ &&H_{\text{QD}}^{N} =\sum_{l=1}^{N}\varepsilon _{l}d_{l}^{\dagger }d_{l}+\sum_{l=1}^{N-1}(V_{l,l+1}\,d_{l}^{\dagger}d_{l+1}+\text{h.c.}),
\nonumber \\
&&H_{\text{QD-QW}} =V_{0}(d_{1}^{\dagger}c_{0}+c_{0}^{\dagger}d_{1}).
\end{eqnarray}

\noindent The operator $c_{i}^{\dagger }$ creates an electron at site $i$, $v$ is the
hopping in the QW, $\varepsilon _{l}$ is the energy level of the dot $l$ and 
$V_{l,l+1}$ is the tunneling coupling between $l\,$th and $(l+1)\,$th QD.
Here $H_{\text{QW}}$ corresponds to the free-particle Hamiltonian on a
lattice with spacing $d$ and whose eigenfunctions are expressed as Bloch
solutions 
\begin{equation}
\left| k\right\rangle =\sum_{j=-\infty }^{\infty }e^{ikdj}\left|
j\right\rangle,
\end{equation}
\noindent where $\left| k\right\rangle $ is the momentum eigenstate and $\left|
j\right\rangle $ is a Wannier state localized at site $j$. The dispersion
relation associated with these Bloch states reads 
\begin{equation}
\varepsilon =2v\cos (kd).
\end{equation}
\noindent Consequently, the Hamiltonian supports an energy band from $-2v$ to $+2v$
and the first Brillouin zone expands the interval $[-\pi /d,\pi /d]$.

The stationary states of the entire Hamiltonian $H$ can be written as 
\begin{equation}
\left| \psi _{k}\right\rangle =\sum_{j=-\infty }^{\infty }a_{j}^{k}\left|
j\right\rangle +\sum_{l=1}^{N}b_{l}^{k}\left| l\right\rangle,
\end{equation}
where the coefficient $a_{j}^{k}$ ($b_{l}^{k}$) is the probability amplitude
to find the electron in the site $j$ of the QW ($l$ of the array) in the
state $k$, namely 
\begin{equation}
a_{j}^{k} =\langle j|\psi _{k}\rangle, \qquad b_{l}^{k} =\langle l|\psi
_{k}\rangle.
\end{equation}

The amplitudes $a_{j}^{k}$ obey the following linear difference equations 
\begin{eqnarray}
\varepsilon a_{j}^{k} &=&v(a_{j-1}^{k}+a_{j+1}^{k})+V_{0}b_{1}^{k}\delta
_{j0},  \nonumber \\
\varepsilon b_{1}^{k} &=&\varepsilon
_{1}b_{1}^{k}+V_{1,2}b_{2}^{k}+V_{0}a_{0}^{k},  \nonumber \\
\varepsilon b_{l}^{k} &=&\varepsilon
_{l}b_{l}^{k}+V_{l,l-1}b_{l-1}^{k}+V_{l,l+1}b_{l+1}^{k},\quad l\neq 1,N, 
\nonumber \\
\varepsilon b_{N}^{k} &=&\varepsilon _{N}b_{N}^{k}+V_{N,N-1}b_{N-1}^{k}.
\end{eqnarray}

\noindent Iterating backwards the equation for $b_{N}^{k}$ we can express the
amplitude $b_{1}^{k}$ in terms of $a_{0}^{k}$ as a continued fraction 
\begin{equation}
b_{1}^{k}=\cfrac{V_{0}a_{0}^{k}}{\varepsilon -\varepsilon _{1}-
\cfrac{V_{1,2}^{2}}{\varepsilon -\varepsilon _{2}-
\mbox{}_{{\displaystyle\ddots} \mbox{\ }_{\displaystyle \varepsilon
-\varepsilon _{N-1} -\cfrac{V_{N-1,N}^{2}}{\varepsilon -\varepsilon _{N}}}}
} }\ .
\end{equation}

\noindent Therefore the equation for $a_{0}^{k}$ can be cast in the form 
\begin{equation}
\varepsilon a_{0}^{k}=v(a_{-1}^{k}+a_{1}^{k})+V_{0}^{2}/Q_{N\ }a_{0}^{k},
\end{equation}
where $Q_{N}$ is the continued fraction 
\begin{equation}
Q_{N}=\varepsilon -\varepsilon _{1}-\cfrac{V_{1,2}^{2}}{\varepsilon
-\varepsilon _{2}- \mbox{}_{{\displaystyle\ddots} \mbox{\ }_{\displaystyle
\varepsilon -\varepsilon _{N-1} -\cfrac{V_{N-1,N}^{2}}{\varepsilon
-\varepsilon _{N}}}} }\ .  \label{continued-fraction}
\end{equation}

In order to study the solutions of the equations (7) we assume that the electrons are described by a plane wave
incident from the far left with unity amplitude and a reflection amplitude $%
r $ and at the far right by a transmission amplitude $t$. Taking this to be
the solution we can write, 
\begin{align}
a_{j}^{k} &=e^{ikdj}+re^{-ikdj}, &j<0, \\
a_{j}^{k} &=te^{ikdj}, &j>1.
\end{align}

The solution of the equations for the $a_{j}^{k}$ '$^{s}$ the can be then obtained iteratively from right to
left. For a given transmission amplitude, the associated incident and
reflection amplitudes may be determined by matching the iterated function to
the proper plane wave at the far left. The transmission probability is given
by $T=\left|t\right|^{2}$ and is obtained from the iterative procedure
described above. In equilibrium we solve the equation for $t$ and $r$ and we
get the following expressions 
\begin{subequations}
\begin{eqnarray}
\label{t-amplitude}
t &=&\frac{2iv\sin (kd)}{2iv\sin (kd)-V_{0}^{2}/Q_{N}}=\frac{Q_{N}}
{Q_{N}+i\Gamma}, \\
r &=&-\frac{V_{0}^{2}/Q_{N}}{2iv\sin (kd)-V_{0}^{2}/Q_{N}}=\frac{i\Gamma}
{Q_{N}+i\Gamma},
\end{eqnarray}
\end{subequations}
\smallskip

\noindent where $\Gamma (\varepsilon )\equiv V_{0}^{2}/2v\sin (kd)$ can be
regarded as the level broadening. Notice that the level broadening can be
fairly well approximated by $\Gamma \simeq V_{0}^{2}/2v$ close to the center
of the band.

The experimentally accessible quantity is the linear conductance $G$ which
is related to the transmission coefficient $T$ at the Fermi energy by the
one-channel Landauer formula at zero temperature 
\begin{eqnarray}
G=\frac{2e^{2}}{h}\,T=\frac{2e^{2}}{h}\,\frac{Q_{N}^{2}}{Q_{N}^{2}+\Gamma^2}.
\end{eqnarray}

It is worth mentioning that the energy levels (zeroes of $Q_{N}$) depend
only on the hopping in the QD array ($V_{n-1,n}$) while $\Gamma $ is only
function of $V_{0}^{2}/v$. Consequently, both magnitudes can be controlled
independently in an actual experiment. This is one of the main advantages of
the present setup.

\section{Results}

\subsection{Short QD array}

Closed expressions for the transmission and reflection coefficients can be
readily obtained when the number of QDs in the array is small. For $N=1$, $%
Q_{1}=\varepsilon -\varepsilon _{1}$ and then we arrive at 
\begin{eqnarray}
T(\varepsilon ) &=&\frac{(\varepsilon -\varepsilon _{1})^{2}}{(\varepsilon
-\varepsilon _{1})^{2}+\Gamma ^{2}},  \nonumber \\
R(\varepsilon ) &=&\frac{\Gamma ^{2}}{(\varepsilon -\varepsilon
_{1})^{2}+\Gamma ^{2}}.
\end{eqnarray}

\noindent The system has an antiresonance at $\varepsilon =\varepsilon _{1}$. The
transmission and the reflection probability are zero and one, respectively.
For $N=2,$ the transmission and reflection coefficients are given by 
\begin{eqnarray}
T(\varepsilon ) &=&\frac{[(\varepsilon -\varepsilon _{1})(\varepsilon
-\varepsilon _{2})-V_{c}^{2}]^{2}}{[(\varepsilon -\varepsilon
_{1})(\varepsilon -\varepsilon _{2})-V_{c}^{2}]^{2}+(\varepsilon
-\varepsilon _{2})^{2}\Gamma ^{2}},  \nonumber \\
R(\varepsilon ) &=&\frac{(\varepsilon -\varepsilon _{2})^{2}\Gamma ^{2}}{%
[(\varepsilon -\varepsilon _{1})(\varepsilon -\varepsilon
_{2})-V_{c}^{2}]^{2}+(\varepsilon -\varepsilon _{2})^{2}\Gamma ^{2}},
\end{eqnarray}

\noindent where $V_{1,2}\equiv V_{c}$. Therefore, the system presents one resonance in 
$\varepsilon -\varepsilon _{2}$ and bonding and antibonding antiresonances
at energies 
\[
\varepsilon =\frac{1}{2}(\varepsilon _{1\text{ }}+\varepsilon _{2\text{ }%
})\pm \frac{1}{2}\sqrt{(\varepsilon _{1\text{ }}-\varepsilon _{2\text{ }%
})^{2}+4V_{c}^{2}}\ . 
\]
The system with $N=3$ side-coupled QDs shows particularly simple solution
for the case $V_{1,2}=V_{2,3}\equiv V_{c}$,

\begin{widetext}

\begin{eqnarray}
T(\varepsilon ) &=&\frac{[(\varepsilon -\varepsilon _{1})(\varepsilon
-\varepsilon _{2})(\varepsilon -\varepsilon _{3})-V_{c}^{2}((\varepsilon
-\varepsilon _{1})+(\varepsilon -\varepsilon _{3}))]^{2}}{[(\varepsilon
-\varepsilon _{1})(\varepsilon -\varepsilon _{2})(\varepsilon -\varepsilon
_{3})-V_{c}^{2}((\varepsilon -\varepsilon _{1})+(\varepsilon -\varepsilon
_{3}))]^{2}+[(\varepsilon -\varepsilon _{2})(\varepsilon -\varepsilon
_{3})-V_{c}^{2}]^{2}\Gamma ^{2}},  \nonumber \\
R(\varepsilon ) &=&\frac{[(\varepsilon -\varepsilon _{2})(\varepsilon
-\varepsilon _{3})-V_{c}^{2}]^{2}\Gamma ^{2}}{[(\varepsilon -\varepsilon
_{1})(\varepsilon -\varepsilon _{2})(\varepsilon -\varepsilon
_{3})-V_{c}^{2}((\varepsilon -\varepsilon _{1})+(\varepsilon -\varepsilon
_{3}))]^{2}+[(\varepsilon -\varepsilon _{2})(\varepsilon -\varepsilon
_{3})-V_{c}^{2}]^{2}\Gamma ^{2}}.
\end{eqnarray}
\end{widetext}

\noindent Clearly for the case with $\varepsilon _{1\text{ }}=\varepsilon _{2\text{ }%
}=\varepsilon _{3},$ the system shows three antiresonances at $\varepsilon
=\varepsilon _{1}$ and $\varepsilon =\varepsilon _{1}\pm \sqrt{2}V_{c}$ and
two resonances at $\varepsilon =\varepsilon _{1}\pm V_{c}$.

\begin{figure}[th]
\centerline{\includegraphics[width=60mm,clip=]{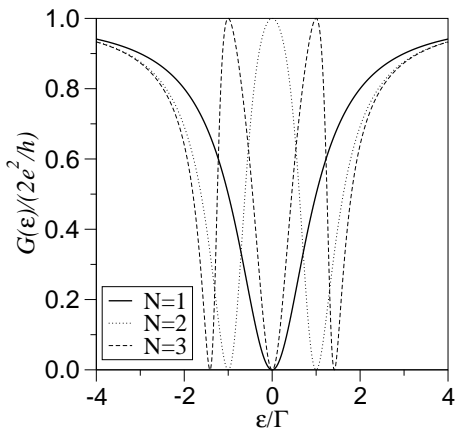}}
\caption{Conductance, in units of $2e^{2}/h$, versus Fermi energy, in units
of the $\Gamma $, for the case of the one, two and three QD array with $%
V_{c}=\Gamma $.}
\label{fig2}
\end{figure}

Figure~\ref{fig2} shows the conductance as a function of the Fermi energy of
the incident electron for $\varepsilon _{i}=0$ ($i=1,\cdots ,N$) and $%
V_{c}=\Gamma $. There exists only one narrow antiresonance in the case of $%
N=1$ QD, and bonding and antibonding antiresonances and one resonance in
zero are clearly revealed for $N=2$. In addition, bonding and antibonding
resonances and zero, bonding and antibonding antiresonances arise when $N=3$.

\subsection{Long QD array}

When the number $N$ of attached QDs is large, we must rely on numerical
calculations. For sake of simplicity we consider a uniform quantum dot array 
$V_{l-1,l}\equiv V_{c}$ and $\varepsilon _{l}=\varepsilon _{0}$. The
continued fraction $Q_{N}$ in Eq.~(\ref{continued-fraction}) is written as $%
Q_{N}=(\varepsilon -\varepsilon _{0})x_{N}$ , where $x_{N}$  satisfies
the following recursive equation, 
\begin{equation}
x_{N}=1-\frac{\alpha }{x_{N-1}},\qquad N=1,2,3,\cdots  \label{map}
\end{equation}
with $x_{1}=1$ and $\alpha \equiv V_{c}^{2}/(\varepsilon -\varepsilon
_{0})^{2}$ for $\varepsilon \neq \varepsilon _{0}$.

\begin{figure}[th]
\centerline{\includegraphics[width=60mm,clip=]{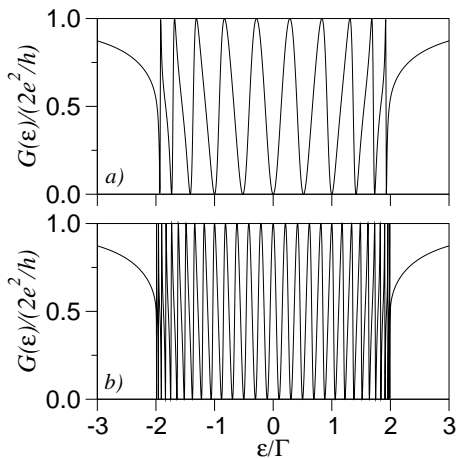}}
\caption{Conductance, in units of $2e^{2}/h$, versus Fermi energy, in units
of $\Gamma $, for a) $N=11$ and b) $N=30$ QD array with $V_{c}=\Gamma $ and $%
\varepsilon _{0}=0$.}
\label{fig3}
\end{figure}

\begin{figure}[ht]
\centerline{\includegraphics[width=60mm,clip=]{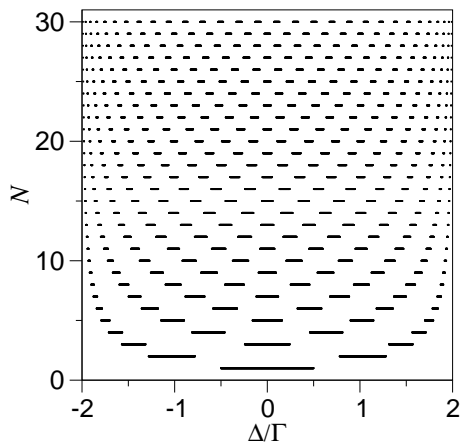}}
\caption{Full width at half minimum of the antiresonances $\Delta$ for
different values of $N$ and $V_{c}=\Gamma$. Each segment joints the two
energies for which conductance becomes $e^2/h$ on every antiresonance.}
\label{fig4}
\end{figure}

For $N$ large the antiresonance appearing in Fig.~\ref{fig2} for $N=1$
evolves as a fast oscillation band. As the number of QDs $N$ increases, this
narrow antiresonance splits into $N$ antiresonances and $N-1$ resonances, as
seen in the upper panel of Fig.~\ref{fig3} for $V_{c}=\Gamma $ and $%
\varepsilon _{0}=0$. On further increasing $N$, the antiresonances never
merge into a single stop-band, as one would naively expect. In fact, it is
not difficult to demonstrate that $Q_{N}=D_{N}/D_{N-1},$ where $D_{N}=\det
(H_{QD}^{N}-\varepsilon I).$ The transmission coefficient can be written as 
\begin{equation}
T=\frac{D_{N}^{2}}{D_{N}^{2}+\Gamma ^{2}D_{N-1}^{2}}.
\end{equation}
Thus, transmission vanishes in the spectrum of $H_{QD}^{N}$ and becomes
unity in the spectrum of the $H_{QD}^{N-1}$. Therefore the conductance show $%
N$ antiresonances and $N-1$ resonances, as we see in Figs.~\ref{fig1} and~%
\ref{fig2}. This statement is further confirmed by plotting the full width
at half minimum of the antiresonances, $\Delta $, as a function of energy,
as shown in Fig.~\ref{fig4} for different values of $N$.

\subsection{Odd-Even Parity}

Here we consider the case when the Fermi energy is pinned at the value of
the energy level of the quantum dot. From equation (7) is straightforward to
prove the existence of an odd-even parity when $\varepsilon _{l}=\varepsilon
_{0},$ 
\begin{align}
G=0,& \qquad N\text{\ odd},  \nonumber \\
G=\frac{2e^{2}}{h},& \qquad N\text{\ even},
\end{align}
as it can be readily checked in Figs.~\ref{fig1} and~\ref{fig2}. This result
also holds for QD arrays with off-diagonal disorder. This odd-even parity is
opposed to the case of an embedded quantum array, where perfect transmission
takes place for odd parity.\cite{oguri,zeng} This property arises from the
fact that energy level of the of the QDs $\varepsilon _{0}$ is always in the
electronic spectrum of the isolated QD array, provided the number of the QDs
is odd. It is straightforward to demonstrate this statement from the
fact $D_{N}$ satisfies the following recursive equation,
\begin{equation}
D_{N}=(\varepsilon -\varepsilon _{0})D_{N-1}-V_{N-1,N}^{2}D_{N-2},\qquad
N=3,4,5\cdots 
\end{equation}
with $D_{1}=\varepsilon -\varepsilon _{0}$ and $D_{2}$ $=(\varepsilon
-\varepsilon _{0})^{2}-V_{1,2}^{2}$. It is clear from Eq. (21) that $D_{N}$ is zero at $%
\varepsilon =\varepsilon _{0}$ if $N$ is odd. Therefore $\varepsilon _{0}$ is
eigenvalue of $H_{QD}^{N}$ and then from equation (19) we obtain that at $%
\varepsilon =\varepsilon _{0}$, $T=0$ (perfect reflection) if $N$ is odd 
and $T=1$ (perfect transmission) for even $N$. 

\subsection{Infinite QD array}

To understand the origin of the fast oscillations of the conductance as a
function of the Fermi energy (Fig.~\ref{fig3}), we now consider the limiting
case $N\to \infty $. Thus, we are faced to a one-dimensional map (\ref{map}%
). This map has two fixed points at \cite{wall} 
\begin{equation}
x_{\pm }^{*}=\frac{1}{2}\,\Big(1\pm \sqrt{1-4\alpha }\Big),
\end{equation}
when $\alpha <1/4$, namely $|\varepsilon -\varepsilon _{0}|>2V_{c}$. The
conductance for $N\to \infty $ is, 
\begin{equation}
G_{\infty }=\frac{2e^{2}}{h}\,\frac{(\left| \varepsilon -\varepsilon
_{0}\right| +\sqrt{(\varepsilon -\varepsilon _{0})^{2}-4V_{c}^{2}})^{2}}{%
(\left| \varepsilon -\varepsilon _{0}\right| +\sqrt{(\varepsilon
-\varepsilon _{0})^{2}-4V_{c}^{2}})^{2}+\Gamma ^{2}},
\end{equation}
for $|\varepsilon |>2V_{c}$. This result explains the smooth tails seen in
Fig.~\ref{fig3} when $|\varepsilon -\varepsilon _{0}|/\Gamma >2$.

The conductance undergoes a bifurcation at $\alpha =1/4$ ($\left|
\varepsilon -\varepsilon _{0}\right| =2V_{c}$), and there are not fixed
points when $\alpha >1/4$, namely $|\varepsilon -\varepsilon _{0}|<2V_{c}$.
Consequently, minute variations of the Fermi energy result in a dramatic
change in the conductance of the QW, as it can be concluded from the lower
panel of Fig.~\ref{fig3}.

\section{Summary}

In this work, we studied the conductance at zero temperature of a side QD
array attached to a QW. For a uniform QD array we found that the system
develops an oscillating band with $N$ antiresonances and $N-1$ resonances
arising from the hybridization of the quasibound levels of the QDs and the
coupling to the QW. The positions of the antiresonances correspond exactly
to the electronic spectrum of the isolated QD array. This property could be
used to measure the energy spectrum of the $N$ QD array. It should be
stressed that the particular setup we suggested allows us to control the
energy and the width of the antiresonances in an independent fashion. When
the number of attached QDs is large, a rich phenomenology appears for
different values of the Fermi energy. When the Fermi energy lies far from
the center of the QW band ($\left| \varepsilon -\varepsilon _{0}\right|
>2V_{c}$), the conductance presents regular and smooth behavior. However,
the conductance strongly fluctuates close to the center of the QW band ($%
\left| \varepsilon -\varepsilon _{0}\right| <2V_{c}$). These results pose a
question about the relevance of the bifurcation at $\left| \varepsilon
-\varepsilon _{0}\right| =2V_{c}$ in actual experiments. Finally, we found
an odd-even parity behavior of the conductance when the Fermi energy lies in
center of the band. If the number of QDs in the array is even, perfect
transmission takes place ($G=2e^{2}/h$). On the contrary, perfect reflection
occurs when this number is odd ($G=0$). This property arises from the
intrinsic electronic properties of the QD array.

\begin{acknowledgments}

 P.\ A.\ O.\ would like to thank financial support from Milenio ICM P99-135-F
and FONDECYT under grant 1020269. Work in Madrid was supported by DGI\-MCyT
(MAT2000-0734) and CAM (07N/0075/2001).

\end{acknowledgments}

\end{document}